\documentclass[a4paper]{article}
\usepackage{epsfig,amssymb}

\def\beq{\begin{eqnarray}}
\def\eeq{\end{eqnarray}}
\def\bsp{\begin{split}}
\def\esp{\end{split}}
\def\lra{\longrightarrow}

\newcommand{\mc}[1]{{\cal #1}}
\input tcilatex
\QQQ{Language}{
British English
}

\begin{document}

\title{\textbf{Magnetic Brane-worlds}}
\author{\textbf{John D. Barrow}\footnote{J.D.Barrow@damtp.cam.ac.uk} ~\textbf{and Sigbj\o rn Hervik}\footnote{S.Hervik@damtp.cam.ac.uk} \\
DAMTP, \\
Centre for Mathematical Sciences,\\
Cambridge University\\
Wilberforce Rd. \\
Cambridge CB3 0WA, UK}
\maketitle

\begin{abstract}
We investigate brane-worlds with a pure magnetic field and a perfect fluid.
We extend earlier work to brane-worlds, and find new properties of the
Bianchi type I brane-world. We find new asymptotic behaviours on approach to
the singularity and classify the critical points of the dynamical phase
space. It is known that the Einstein equations for the magnetic Bianchi type
I models are in general oscillatory and are believed to be chaotic, but in
the brane-world model this chaotic behaviour does not seem to be possible.
\end{abstract}

\section{Introduction}

Developments in string theory have inspired the construction of
brane-worlds, in which standard-model gauge fields are confined to our
three-dimensional brane-world, while gravity propagates in all the spatial
dimensions. A simple five-dimensional class of such models allows for a
non-compact extra dimension via a novel mechanism for localization of
gravity around the brane at low energies\cite{Randall:1999vf,Randall:1999ee}%
. This mechanism is the warping of the metric by a negative five-dimensional
cosmological constant. These models have been generalized to admit
cosmological branes~\cite{sms}, and they provide an interesting arena in
which to impose cosmological tests on extra-dimensional generalizations of
Einstein's theory. One of the unusual features of this structure is that
brane-worlds can feel the effects of non-local anisotropic stresses in the
bulk dimensions. These graviton stresses are unconstrained by dynamical
equations specified on the brane but for a wide range of behaviours they
completely determine the evolution of small anisotropies on the brane-world
and can slow the their decay to such an extent that their effects are
directly detectable today in the temperature anisotropy of the microwave
background radiation\cite{bm}. The evolution of anisotropic brane-worlds has
been studied recently by several authors \cite
{Campos:2001pa,Toporensky:2001hi,Santos:2001nt,Campos:2000cn,Chen:2001wg,MSS:2001}.
They consider only the behaviour of simple anisotropies in universes with
perfect fluid matter sources. However, it is known that the evolution of
simple anisotropic universes with isotropic three-curvature is very
sensitive to the presence of trace-free matter sources with anisotropic
pressures, \cite{Barrow:1997sy,jb,bm2}. Such stresses are inevitable in
anisotropic cosmologies at early times because of the presence of
collisionless gravitons, collisionless asymptotically-free particles, and
electric or magnetic fields. These stresses also mimic the effects of more
general anisotropic curvature anisotropies in more general anisotropic
universe (because the latter can be viewed as an effective stress of
long-wavelength gravitational waves). Small anisotropic pressures serve to
slow the decay of the shear anisotropy to a logarithm of the time during the
radiation era and to a slow power law during the dust era. These anisotropic
stresses can arise because of intrinsic anisotropic stresses on the brane or
via the induced graviton stresses from the bulk. In this paper we are going
to consider the detailed behaviour of the simplest example: a pure magnetic
field on a flat anisotropic brane with Bianchi type I geometry. Whereas the
study of ref. \cite{bm} was interested in the late-time behaviour we shall
be interested primarily in the behaviour as $t\rightarrow 0.$

The possibility of isotropization and homogenisation of cosmological models
have been an important study since the original equations for gravitational
interaction where put down in 1915 by Einstein. The general relativistic
anisotropic models called the Bianchi models have been studied in great
detail both in general relativistic cosmologies \cite
{Ruban,Barrow:1997sy,Hervik:2000ed} and to some extent in the brane-world
scenario . A feature that has been known to exist in the Bianchi models are
oscillatory solutions\cite{LeBlanc,Spokoiny:1981fs}, and even chaotic
solutions \cite{Belinsky:1970ew,jb2,jb3,cb,dem,DH2000}. The chaotic
behaviour in the general relativistic vacuum Bianchi type VIII and IX models
has been well studied, and chaotic behaviour for the general relativistic
Bianchi type I model with a magnetic field has been conjectured \cite
{LeBlanc} and studied for the case of form fields \cite{dh2}. There are many
similarities between magnetic Bianchi universes and the behaviour of
pre-big-bang string cosmologies \cite{bk}. For further discussion of
magnetic field effects in general cosmological models see also refs. \cite
{tsag1,tsag2}.

In the brane-worlds model the dynamics change because of the influence of
terms varying as the square of the isotropic and anisotropic pressure as $%
t\rightarrow 0$. We investigate the Bianchi type I brane-world with a
perfect fluid and a pure magnetic field in the context of a possible chaotic
behaviour. This extends the general relativistic study by LeBlanc \cite
{LeBlanc} to the brane-world scenario. We will show that unlike the general
relativity case, chaotic behaviour is probably completely absent for all
perfect fluids as a result of the non-linear stresses.

This paper is organized as follows: First we put down the fundamental
equations of motion. We use the representation of LeBlanc to facilitate
comparisons, and introduce expansion-normalized variables. Since we are
primarily interested in the initial singularity, we find some approximate
solutions near the initial singularity for different magnetic
configurations. All of these approximate solutions correspond to equilibrium
points for the full set of equations, and so we provide a stability analysis
of the equilibrium points found.

\section{Equations of motion for a Bianchi type I brane-world}

The Bianchi cosmological models can be classified according to the
commutation relations of their invariant 1-forms, $\omega ^i$. We will use
an orthonormal frame of vector fields $\{e_\mu \}$, and align the frame so
that $e_0=\partial _t$, the fluid vector field. This implies that the
spatial triad $\{e_a\}$ spans the tangent space at each point of the group
orbits. Using the commutation relations for this vector basis, the
commutator functions $C_{~\beta \gamma }^\alpha (t)$ are given by 
\begin{eqnarray}
\lbrack e_\alpha ,e_\beta ]=C_{~\alpha \beta }^\gamma (t)e_\gamma 
\end{eqnarray}

In the Bianchi Class A case we have the relations\footnote{%
Latin indices have the range 1-3, while Greek indices have the range 0-3.} 
\begin{eqnarray}
C_{~0b}^a(t)=-\Theta _{~b}^a(t)-\epsilon _{~bc}^a\Omega
^c(t),~C_{~bc}^a=\epsilon _{bcd}n^{ad}
\end{eqnarray}
The $\Theta _{~b}^a(t)$ are the frame components of the volume expansion
tensor of the fluid, or the extrinsic curvature. The vector $\Omega ^c(t)$
is the angular velocity of the spatial triad with respect to a set of
Fermi-propagated axes. The extrinsic curvature $\Theta _{\mu \nu }$ is
separated into its trace, $\Theta $, the volume expansion, and its
trace-free part $\sigma _{\mu \nu },$ the shear: 
\begin{eqnarray}
\Theta _{\mu \nu }=\frac 13\Theta h_{\mu \nu }+\sigma _{\mu \nu }
\end{eqnarray}
We will also assume that we have a non-zero electromagnetic (EM) field and a
non-interacting perfect fluid. The perfect-fluid energy momentum tensor is 
\begin{eqnarray}
T_{\mu \nu }^{PF}=(\rho +p)u_\mu u_\nu +pg_{\mu \nu }
\end{eqnarray}
Further, we will assume a barotropic equation of state for the fluid for the
pressure $p$ and density $\rho $: 
\begin{eqnarray}
p=(\gamma -1)\rho 
\end{eqnarray}
where $\gamma \in [0,2]$ is constant. The perfect fluid solves the usual
conservation equation 
\begin{eqnarray}
\dot{\rho}+\gamma \Theta \rho =0  \label{con}
\end{eqnarray}

The energy-momentum tensor of the EM field can be expressed as 
\begin{eqnarray}
T_{\mu \nu }^{EM}=\rho _{EM}u_\mu u_\nu +p_{EM}h_{\mu \nu }+\pi _{\mu \nu },
\end{eqnarray}
where $\rho _{EM}$ and $p_{EM}$ are the energy density and the isotropic
pressure, respectively, and the tensor $h_{\mu \nu }=g_{\mu \nu }+u_\mu
u_\nu $ projects orthogonal to the $u^\mu $ on the brane, and $\pi _{\mu \nu
}$ is the tracefree anisotropic pressure tensor. In the case of an electric
field $E$ and magnetic field $B$ the energy-momentum tensor can be written
in general as 
\begin{eqnarray}
T_{\mu \nu }^{EM}=-\frac 14F^{\alpha \beta }F_{\alpha \beta }g_{\mu \nu
}+F_{\mu \alpha }F_\nu ^{~\alpha }
\end{eqnarray}
where the anisotropic stress is 
\begin{eqnarray}
\pi _{\mu \nu }=-E_\mu E_\nu -B_\mu B_\nu +\frac 13(E^2+B^2)h_{\mu \nu }
\end{eqnarray}
and the pressure and density are given by 
\begin{eqnarray}
\rho _{EM}=3p_{EM}=\frac 12(E^2+B^2)
\end{eqnarray}
Demanding that the perfect fluid and the EM field be separately conserved
requires eqn. (\ref{con}) 
\begin{eqnarray}
\dot{\rho}_{EM}+\Theta (\rho _{EM}+p_{EM})+\sigma ^{\mu \nu }\pi _{\mu \nu
}=0.
\end{eqnarray}

For a Bianchi type I model the commutator between the vectors of the spatial
triad all commute so 
\begin{eqnarray}
C_{~bc}^a=0
\end{eqnarray}
This allows us to have a pure magnetic field\cite{Hughston}. This is not the
case for all Bianchi models. We shall therefore specialise to the case of a
Bianchi type I model with a vanishing electric field.

We now consider the evolution equations on the brane \cite{Maartens:2001jx,
Maartens:2000fg} for the expansion scale factor, shear, and energy
densities. The latter consist of the perfect fluid density, the magnetic
energy and the dark energy term $\mathcal{U}$. The non-local energy
conservation equation for the dark energy $\mathcal{U}$ is 
\begin{eqnarray}
&&\dot{\mathcal{U}}+\frac 43\Theta \mathcal{U}+\sigma ^{\mu \nu }\mathcal{P}%
_{\mu \nu }=  \nonumber \\
&&\frac{\kappa ^4}{12}\left[ 3\pi ^{\mu \nu }\dot{\pi}_{\mu \nu }+3(\rho
+p)\sigma ^{\mu \nu }\pi _{\mu \nu }+\Theta \pi ^{\mu \nu }\pi _{\mu \nu
}-\sigma ^{\mu \nu }\pi _{\alpha \mu }\pi _\nu ^{~\alpha }\right]
\end{eqnarray}
where the overdot means $u^\mu \nabla _\mu \equiv d/dt$, $\mathcal{P}_{\mu
\nu }$ is the unconstrained bulk graviton stress, and $\kappa ^2\equiv 8\pi
G $. The constraint equation is 
\begin{eqnarray}
H^2=\frac \Lambda 3+\frac 16\sigma ^{\mu \nu }\sigma _{\mu \nu }-\frac 16%
\mathcal{R}+\frac{\kappa ^2}3\left[ \rho \left( 1+\frac \rho {2\lambda
}\right) -\frac 3{4\lambda }\pi _{\mu \nu }\pi ^{\mu \nu }\right] +\frac{2%
\mathcal{U}}{\kappa ^2\lambda }  \label{constraint}
\end{eqnarray}
where $\mathcal{R}$ is the 3-curvature of the brane, $\lambda $ is the brane
tension, $\Lambda $ the cosmological constant, and $H=\frac 13\Theta $ is
the Hubble expansion parameter. The Gauss-Codazzi shear propagation
equations are: 
\begin{eqnarray}
\dot{\sigma}_{<\mu \nu >}+\Theta \sigma _{\mu \nu }=\kappa ^2\pi _{\mu \nu }+%
\frac{\kappa ^2}{2\lambda }\left[ -(\rho +3p)\pi _{\mu \nu }+\pi _{\alpha
<\mu }\pi _{~\nu >}^\alpha \right] +\frac 6{\kappa ^2\lambda }\mathcal{P}%
_{\mu \nu }-\mathcal{R}_{<\mu \nu >}  \label{CMeq}
\end{eqnarray}
where $A_{<\mu \nu >}$ for a tensor $A_{\mu \nu }$, means the projected,
symmetric and trace-free part \cite{Maartens:2001jx, Maartens:2000fg}. The
volume expansion propagation is governed by Raychaudhuri's equation 
\begin{eqnarray}
&&\dot{\Theta}+\frac 13\Theta ^2+\sigma ^{\mu \nu }\sigma _{\mu \nu }+\frac
12\kappa ^2(\rho +3p)-\Lambda =  \nonumber \\
&&-\frac 1{2\lambda \kappa ^2}\left[ \kappa ^4(2\rho ^2+3\rho p)+12\mathcal{U%
}\right]
\end{eqnarray}
Here, for simplicity we shall assume that $\mathcal{P}_{\mu \nu }=0$ in
order for the equations to close and we set $\mathcal{R}_{<\mu \nu >}=0$ in
the Bianchi I universe and also assume $\Lambda =0$. The dark energy, $%
\mathcal{U}$ (which can in principle be of either sign) on the other hand
cannot be set to zero in the presence of a non-zero anisotropic stress. In
addition to these equations the magnetic field satisfies Maxwell's
equations: 
\begin{eqnarray}
\dot{B}_\mu =-\frac 23\Theta B_\mu +\sigma _{\mu \nu }B^\nu
\end{eqnarray}
Since we shall be concerned with the early time behaviour we can ask if
there is an initial singularity. In the brane-world we can write the
energy-momentum tensor as a sum of the classical energy-momentum tensor and
a term that come from brane-effects only. A sufficient condition for a
singularity is that the classical energy-momentum tensor obeys the strong
energy condition (SEC), \emph{and} that $\mathcal{U}>0$. We must however
emphasize that we can still have a singularity even if this condition is
violated.

The equations in the orthonormal frame reduce to the following first-order
system: 
\begin{eqnarray}
\dot{\Theta} &=&\partial _t\Theta  \\
\dot{B}_a &=&\partial _tB_a-\epsilon _{abc}\Omega ^cB^b \\
\dot{\sigma}_{ab} &=&\partial _t\sigma _{ab}+2\sigma _{(a}^d\epsilon
_{b)cd}\Omega ^c \\
\dot{\pi}_{ab} &=&\partial _t\pi _{ab}+2\pi _{(a}^d\epsilon _{b)cd}\Omega ^c
\end{eqnarray}
We still have an unused gauge freedom to choose the orientation of the
spatial triad. We choose the orientation so that the magnetic field is
aligned with $e_1$, thus ensuring $B_2=B_3=0$ while $B_1=h$. From the
Maxwell equations for $B_a$ this yields the constraints 
\begin{eqnarray}
\Omega ^2=-\sigma _{13},~\Omega ^3=\sigma _{12}
\end{eqnarray}
The $\Omega ^1$ is still left undetermined, but to simplify the equations we
will choose a frame so that $\Omega ^1=0$. This choice of gauge makes the
anisotropic-stress tensor diagonal (as in the general case of refs.\cite
{Barrow:1997sy}, \cite{bm2}), with 
\begin{eqnarray}
\pi _{\mu \nu }=D_{\mu \nu }h^2
\end{eqnarray}
where $D_{\mu \nu }=$ $diag(0,-\frac 23,\frac 13,\frac 13)$.

Introducing the parameters $\sigma _{\pm }$, defined by

\[
(\sigma _{11},\sigma _{22},\sigma _{33})=(-2\sigma _{+},\sigma _{+}+\sqrt{3}%
\sigma _{-},\sigma _{+}-\sqrt{3}\sigma _{-}),
\]
allows the full set of equations to be written as a first-order dynamical
system: 
\begin{eqnarray}
\partial _th &=&-\frac 23\Theta h-2\sigma _{+}h \\
\partial _t\rho  &=&-\gamma \Theta \rho  \\
\partial _t\Theta  &=&-\frac 13\Theta ^2\ -\sigma ^2-\frac 12[(3\gamma
-2)\rho +h^2]  \nonumber \\
&&-\frac \chi 8[4(3\gamma -1)\rho ^2+(6\gamma +4)\rho h^2+3h^4]-6\chi 
\mathcal{U} \\
\partial _t\mathcal{U} &=&-\frac 43\Theta \mathcal{U}-\frac{h^2}6\left[
\Theta h^2+\frac 53\sigma _{+}h^2-3\gamma \sigma _{+}\rho \right]  \\
\partial _t\sigma _{+} &=&-\Theta \sigma _{+}-\sigma _{12}^2-\sigma
_{13}^2+\frac 13h^2\left( 1-\frac \chi 2\left[ (3\gamma -2)\rho +\frac
43h^2\right] \right)  \\
\partial _t\sigma _{-} &=&-\Theta \sigma _{-}-\frac 1{\sqrt{3}}(\sigma
_{12}^2-\sigma _{13}^2) \\
\partial _t\sigma _{23} &=&-\Theta \sigma _{23}-2\sigma _{12}\sigma _{13} \\
\partial _t\sigma _{12} &=&-\Theta \sigma _{12}+(3\sigma _{+}+\sqrt{3}\sigma
_{-})\sigma _{12}+\sigma _{13}\sigma _{23} \\
\partial _t\sigma _{13} &=&-\Theta \sigma _{13}+(3\sigma _{+}-\sqrt{3}\sigma
_{-})\sigma _{13}+\sigma _{12}\sigma _{23},
\end{eqnarray}
where $\chi =(\kappa ^2\lambda )^{-1}$ is constant and where we have set $%
\kappa =1$ implicitly by making the redefinitions: $\kappa h\longrightarrow h
$ and $\kappa ^2\rho \longrightarrow \rho $. Interestingly, there is a
further symmetry in these equations. By a change of variables 
\begin{eqnarray}
\sigma _{12}+i\sigma _{13} &=&\sigma _Ae^{i\phi } \\
\sqrt{3}\sigma _{-}+i\sigma _{23} &=&(\sigma _B+i\sigma _C)e^{2i\phi },
\end{eqnarray}
the equation for $\phi $ turns into 
\begin{eqnarray}
\partial _t\phi =\sigma _C
\end{eqnarray}
and can be completely eliminated from the system. Thus, this variable can be
ignored. The remaining equations for the shear are 
\begin{eqnarray}
\partial _t\sigma _{+} &=&-\Theta \sigma _{+}-\sigma _A^2+\frac 13h^2\left(
1-\frac \chi 2\left[ (3\gamma -2)\rho +\frac 43h^2\right] \right)  \\
\partial _t\sigma _A &=&(-\Theta +3\sigma _{+}+\sigma _B)\sigma _A \\
\partial _t\sigma _B &=&-\Theta \sigma _B-\sigma _A^2+2\sigma _C^2 \\
\partial _t\sigma _C &=&(-\Theta -2\sigma _B)\sigma _C
\end{eqnarray}

A further simplification is achieved by defining a new set of expansion
normalized variables: 
\begin{eqnarray}
\Sigma _{+} &=&\frac{3\sigma _{+}}\Theta  \\
\Sigma _{A,B,C} &=&\frac{\sqrt{3}\sigma _{A,B,C}}\Theta  \\
U &=&\frac{\mathcal{U}}{\Theta ^2} \\
\mathcal{H} &=&\frac h\Theta  \\
\Omega  &=&\frac{3\rho }{\Theta ^2}
\end{eqnarray}
We introduce a new time variable by 
\begin{eqnarray}
\frac{dt}{d\tau }=\frac 3\Theta   \label{timeshift}
\end{eqnarray}
and denote differentiation with respect to $\tau $ by $^{\prime },$ so the
evolution equations become: 
\begin{eqnarray}
\Theta ^{\prime } &=&-(1+q)\Theta  \\
U^{\prime } &=&(2q-2)U-\frac{\mathcal{H}^2}{18}\left( 9\mathcal{H}^2+5\Sigma
_{+}\mathcal{H}^2-3\gamma \Omega \right) \Theta ^2 \\
\Omega ^{\prime } &=&[2q-(3\gamma -2)]\Omega  \\
\mathcal{H}^{\prime } &=&\left( q-1-2\Sigma _{+}\right) \mathcal{H} \\
\Sigma _{+}^{\prime } &=&(q-2)\Sigma _{+}-3\Sigma _A^2+3\mathcal{H}^2\left(
1-\frac \chi 6\left[ (3\gamma -2)\Omega +{4}\mathcal{H}^2\right] \Theta
^2\right)  \\
\Sigma _A^{\prime } &=&\left( q-2+3\Sigma _{+}+\sqrt{3}\Sigma _B\right)
\Sigma _A \\
\Sigma _B^{\prime } &=&(q-2)\Sigma _B+2\sqrt{3}\Sigma _C^2-\sqrt{3}\Sigma
_A^2 \\
\Sigma _C^{\prime } &=&\left( q-2-2\sqrt{3}\Sigma _B\right) \Sigma _C
\end{eqnarray}
where 
\begin{eqnarray}
q &\equiv &1+\Sigma +\frac 12(3\gamma -4)\Omega   \nonumber \\
&&+\chi \left[ (3\gamma -2)\frac{\Omega ^2}6+\frac 34\gamma \Omega \mathcal{H%
}^2+\frac 94\mathcal{H}^4\right] \Theta ^2 \\
\Sigma  &\equiv &\Sigma _{+}^2+\Sigma _A^2+\Sigma _B^2+\Sigma _C^2 \\
1 &=&\Sigma +\Omega +\frac 32\mathcal{H}^2+\chi \left[ \frac{\Omega ^2}%
6+\frac \Omega 2\mathcal{H}^2-\frac 98\mathcal{H}^4\right] \Theta ^2+18\chi U
\end{eqnarray}
and the last of these is the constraint equation. We note that in the limit
of general relativity ($\chi =0$), the variables $U$ and $\Theta $ decouple
and $\Omega $ can then be determined from the constraint equation, leaving
us with 5 fundamental variables: $(\mathcal{H},\Sigma _{+},\Sigma _{A,B,C})$.

\subsection{Flat Friedmann universe}

Let us first look at one of the simplest cases, the flat isotropic Friedmann
brane-world. We set $\mathcal{H}=\Sigma _{+}=\Sigma _{A,B,C}=0$, and the
solution can now be derived exactly. The result is a two parameter family $%
(\Theta _0,K)$ of exact solutions given by: 
\begin{eqnarray}
\Omega (\tau ) &=&\left( \frac \chi 6\Theta _0^2e^{-3\gamma \tau
}+1+Ke^{(3\gamma -4)\tau }\right) ^{-1} \\
18\chi U(\tau ) &=&Ke^{(3\gamma -4)\tau }\Omega (\tau ) \\
\Theta (\tau ) &=&\Theta _0e^{-\frac 32\gamma \tau }\Omega (\tau )^{-\frac
12}
\end{eqnarray}

We note that if $\gamma <\frac 23$ the Friedmann brane-universe will be $U$
dominated near the initial singularity. Otherwise, the quadratic matter-term
will dominate initially in eqn. (\ref{constraint}). This is quite different
from the general relativistic case. In the general relativity case ($\chi =0$%
), the $\Omega $ term will dominate initially in an isotropic universe,
where $\Omega =1$ exactly. The metric for a flat Friedmann brane-world can
now be written as 
\begin{eqnarray}
ds^2=-\frac 9{\Theta (\tau )^2}d\tau ^2+e^{{2}\tau }(dx^2+dy^2+dz^2)
\end{eqnarray}
By integrating eqn. (\ref{timeshift}) we can restore the comoving proper
time coordinate $t$.

\subsection{Non-magnetic Bianchi type I solutions}

Now we examine the invariant solution subspace $\mathcal{H}=\Sigma _{A,C}=0$
of Bianchi type I universes containing perfect fluid and magnetic field. If
we introduce an angular variable $\phi $ and parameters $A,K$ and $\Theta _0$
the solutions can be written in parametric form: 
\begin{eqnarray}
\Omega (\tau ) &=&\left( \frac \chi 6\Theta _0^2e^{-3\gamma \tau
}+1+Ke^{(3\gamma -4)\tau }+A^2e^{3(\gamma -2)\tau }\right) ^{-1} \\
18\chi U(\tau ) &=&Ke^{(3\gamma -4)\tau }\Omega (\tau ) \\
\Theta (\tau ) &=&\Theta _0e^{-\frac 32\gamma \tau }\Omega (\tau )^{-\frac
12} \\
\Sigma _{+}(\tau ) &=&Ae^{\frac 32(\gamma -2)\tau }\Omega (\tau )^{\frac
12}\cos \phi \\
\Sigma _B(\tau ) &=&Ae^{\frac 32(\gamma -2)\tau }\Omega (\tau )^{\frac
12}\sin \phi .
\end{eqnarray}

These solutions have been found earlier \cite{Chen:2001wg}. Let us look at
the initial singularity in this case, which is approached as $\tau
\longrightarrow -\infty $. The behaviour near the singularity in this case
depends strongly on the parameter $\gamma $~\cite{Toporensky:2001hi}. If the
perfect fluid has $\gamma <1$ then the isotropic perfect fluid stresses are
dominated by the shear anisotropy at early times and a Kasner-like
singularity results (in the general relativistic case this occurs for $%
\gamma <2$ ), as $\tau \longrightarrow -\infty $: 
\begin{eqnarray}
\Omega  &\approx &A^{-2}e^{{3}(2-\gamma )\tau } \\
\Sigma _{+} &\approx &\cos \phi  \\
\Sigma _B &\approx &\sin \phi .
\end{eqnarray}
In this case, the behaviour is that of a vacuum Kasner universe, and the
metric near the initial singularity can be approximated by the Kasner
metric, 
\begin{eqnarray}
ds^2=-dt^2+t^{2p_1}dx^2+t^{2p_2}dy^2+t^{2p_3}dz^2,
\end{eqnarray}
where $\sum p_i=\sum p_i^2=1$. In the case of $\gamma =1$, the solutions can
be written similar to the Kasner metric, but now the exponents fulfil the
equations $\sum p_i=\cos \psi \geq 0$, $\sum p_i^2=\cos ^2\psi $. The
solution space is thus as a half-sphere and the situation is analogous to
that of general relativity when $\gamma =2$.

If $\gamma >1,$ we get a matter-dominated behaviour as $\tau \longrightarrow
-\infty $ (this type of behaviour cannot occur in general relativity because
it would require $\gamma >2$) 
\begin{eqnarray}
\Omega  &\approx &\frac 6{\chi \Theta _0^2}e^{3\gamma \tau } \\
\Sigma _{+} &\approx &\frac{\sqrt{6}A}{\sqrt{\chi }\Theta _0}e^{(3\gamma
-1)\tau }\cos \phi  \\
\Sigma _B &\approx &\frac{\sqrt{6}A}{\sqrt{\chi }\Theta _0}e^{(3\gamma
-1)\tau }\sin \phi 
\end{eqnarray}
In this case we see that the behaviour is quasi-isotropic. The shear is
going exponentially towards zero in terms of $\tau $. The metric in this
case can be approximated by 
\begin{eqnarray}
ds^2=-dt^2+t^{\frac 2{3\gamma }}\left( e^{-\frac 43\omega (t)\cos (\phi
)}dx^2+e^{\frac 43\omega (t)\cos \left( \phi +\frac \pi 3\right)
}dy^2+e^{\frac 43\omega (t)\cos \left( \phi -\frac \pi 3\right) }dz^2\right) 
\end{eqnarray}
where 
\begin{eqnarray}
\omega (t)\equiv \frac{A\gamma }{(3\gamma -1)}\left( \Theta _0^2\gamma \sqrt{%
\frac \chi 6}\right) ^{-\frac 1{3\gamma }}t^{1-\frac 1{3\gamma }}
\end{eqnarray}
This metric differs slightly from the Friedmann metric, since the quadratic
matter term dominates in the initial singularity. In some sense it
asymptotes to a model with zero brane tension. It is a
non-general-relativistic model which was first investigated by Bin\`{e}truy,
Deffayet and Langlois \cite{BDL}.

\section{The singularity in magnetic Bianchi type I brane-worlds}

We will now investigate a more general solution of the magnetic Bianchi type
I brane-world. We have just seen that the quadratic density terms affect the
singularity significantly in the non-magnetic case whenever $\gamma >1.$
Therefore we expect that the anisotropic tracefree stresses contributed by
the magnetic field will have a significant new anisotropising effect on the
form of the solution near the singularity in many cases where $\gamma \geq
4/3$. It is useful to compare with the general relativistic solutions with a
magnetic field. If we set $\chi =U=\Sigma _{A,C}=\Omega =0$ we get the
so-called Rosen solutions. These are 2-parameter pure magnetic solutions,
and an interesting point about these solutions is that the near the initial
singularity the behaviour is that of a Kasner vacuum. The solutions can is
mapped onto the straight line $\Sigma _B=C(\Sigma _{+}-2)$ where we have the
bounds $\Sigma _{+}^2+\Sigma _B^2\leq 1,~3C^2\leq 1$.

\subsection{Rosen branes}

Let us try to find similar solutions for the brane-world scenario. First we
look at the invariant space $\Omega =0=\Sigma _{A,C}$. We make the ansatz 
\begin{eqnarray}
\mathcal{H}^{\prime }=\tilde{p}\mathcal{H}
\end{eqnarray}
where $\tilde{p}$ is a positive constant. Clearly $\mathcal{H}\propto e^{%
\tilde{p}\tau }$, thus $\mathcal{H}$ is exponentially decreasing towards the
initial singularity $\tau \longrightarrow -\infty $. Let us assume that we
are in the vicinity of the Kasner circle, i.e. $\Sigma _{+}$ and $\Sigma _B$
are approximately constants and $q\approx 2$. For this to hold $\mathcal{H}%
^2\Theta $ must also be exponentially decreasing towards the initial
singularity. Since $\Theta ^{\prime }=-(1+q)\Theta $, we get the inequality: 
\begin{eqnarray}
q-3-4\Sigma _{+}>0~\Rightarrow ~-\Sigma _{+}>\frac 14
\end{eqnarray}
In addition, we get $\Sigma _{+}^2+\Sigma _B^2=1$ and near the initial
singularity there are solutions of the form: 
\begin{eqnarray}
\Theta  &=&\Theta _0e^{-3\tau } \\
\mathcal{H} &=&\mathcal{H}_0e^{(1-2\cos \phi )\tau } \\
\Sigma _{+} &=&\cos \phi  \\
\Sigma _B &=&\sin \phi ,
\end{eqnarray}
where the constant $\phi $ has to satisfy 
\begin{eqnarray}
\cos \phi <-\frac 14.
\end{eqnarray}
The function $U$ is 
\begin{eqnarray}
U=\left\{ 
\begin{array}{r@{\quad;\quad}l}
U_0e^{2\tau }+\frac{\mathcal{H}_0^4\Theta _0^2}{72}\cdot \left( \frac{%
9+5\cos \phi }{1+2\cos \phi }\right) \cdot e^{-2(4\cos \phi +1)\tau } & %
\hspace{0.3in}2\cos \phi \neq -1 \\ 
\left( U_0-\frac{13\mathcal{H}_0^4\Theta _0^2}{36}\tau \right) e^{2\tau } & %
\hspace{0.3in}2\cos \phi =-1
\end{array}
\right. 
\end{eqnarray}

\subsection{Shear-dominated singularity with perfect fluid}

Let us again assume that the shear dominates near the initial singularity,
but now we insert a perfect fluid into the equations, so we will only assume
that $\Sigma _{A,C}=0$. We will also assume that we are approximately on the
Kasner circle, i.e. $\Sigma =1$. Thus we get $q=2$ and $\Sigma ^{\prime }=0$%
. The same reasoning as before yields the bound 
\begin{eqnarray}
-\Sigma _{+}>\frac 14  \label{bound1}
\end{eqnarray}
In addition, from the $\mathcal{H}^2\Omega \Theta ^2$ term, we get the bound 
\begin{eqnarray}
-\Sigma _{+}>\frac 14(3\gamma -2).  \label{bound2}
\end{eqnarray}
We note that for $\gamma <1$ the bound (\ref{bound1}) is sufficient for
bound (\ref{bound2}) to hold. In the case $\gamma >1$ the bound (\ref{bound2}%
) is sufficient for bound (\ref{bound1}) to hold. The behaviour for $\Theta
,~\mathcal{H},~\Sigma _{+}$ and $\Sigma _B$ is the same to lowest order as
in the previous example, except that $\Sigma _{+}$ has to satisfy both (\ref
{bound1}) and (\ref{bound2}). The evolution for $\Omega $ is to the lowest
order 
\begin{eqnarray}
\Omega =\Omega _0e^{3(2-\gamma )\tau }
\end{eqnarray}

\subsection{Fluid-dominated singularity}

As we saw in the simple case with no magnetic field present, perfect-fluid
matter can dominate the singularity if $\gamma >1$. This is also the case
with a magnetic field. In that case we have the term $\frac \chi 6\Omega
^2\Theta ^2\approx 1$. Let us assume that this is the case here. For this to
happen we clearly require 
\begin{eqnarray}
q=3\gamma -1.
\end{eqnarray}
On the other hand, the shear has to be exponentially decreasing in $\tau $
as we go towards the initial singularity, so 
\begin{eqnarray}
q-2>0~\Rightarrow ~\gamma >1
\end{eqnarray}
But from the $\mathcal{H}^2\Omega \Theta ^2$ term, we get an even stronger
bound: 
\begin{eqnarray}
\gamma >\frac 43
\end{eqnarray}
The physical explanation for this bound in order that the solutions to be
stable in the past is evident: if the perfect fluid dominates the past
singularity, it inevitably drives the universe towards isotropy. Near
isotropy the magnetic field behaves as a $\gamma _{EM}=\frac 43$ perfect
fluid, thus if $\gamma <\frac 43$, the magnetic field will become dominant
in the past. The fluid-dominated universe will therefore be unstable in the
past for $\gamma <\frac 43$. If the inequality, $\gamma >\frac 43$, is
fulfilled then the evolution near the initial singularity is approximated by 
\begin{eqnarray}
\Theta  &=&\Theta _0e^{-3\gamma \tau }, \\
\Omega  &=&\frac{\sqrt{6}}{\sqrt{\chi }\Theta _0}e^{3\gamma \tau }, \\
\mathcal{H} &=&\mathcal{H}_0e^{(3\gamma -2)\tau }, \\
\Sigma  &=&\Sigma _0e^{6(\gamma -1)\tau }.
\end{eqnarray}

\subsection{$U$-dominated and $\mathcal{H}$-dominated singularity}

Let us look at the other terms in the constraint equation and check whether
the dark energy, $U,$ or the magnetic field, $\mathcal{H}$, can be dominant
near the initial singularity. As we have already seen, the $U$ term
dominates in the Friedmann case if $\gamma <\frac 23$. This is the only
situation where the $U$ term is allowed to dominate. For the $\sqrt{\chi }%
\mathcal{H}^2\Theta \equiv H$ term, the constraint equation allows this term
to grow to arbitrary large values. Let us consider the case where $H$
diverges. The equation for $H$ is 
\begin{eqnarray}
H^{\prime }=(q-3-4\Sigma _{+})H.
\end{eqnarray}
For $H$ to diverge at the initial singularity, $(q-3-4\Sigma _{+})<0$. This
is not possible if $H$ is large. Thus $H$ cannot diverge in the initial
singularity.

We note however that there are solutions of the form 
\begin{eqnarray}
H^2 &\propto &\frac 1{K-q_0\tau } \\
\Sigma _{+} &\propto &\frac 1{K-q_0\tau }
\end{eqnarray}
but these do not describe a physical singularity. These are merely a result
of the choice of time coordinate. These solutions yield 
\begin{eqnarray}
\Theta \propto K-q_0\tau ,
\end{eqnarray}
and are only turning points in the evolution of the brane, where the brane
recollapses. The infinity in $H$ in the phase space does not represent a
physical singularity in the brane evolution.

There are also solutions where $H$ is initially a constant. One of these has
a particular value of $q$: 
\begin{eqnarray}
q=\frac 1{22}(89+\sqrt{3345})
\end{eqnarray}
The solutions in this case are 
\begin{eqnarray}
\Sigma _{+} &=&\frac 1{88}(23+\sqrt{3345}) \\
\Theta  &=&\Theta _0e^{-4(1+\Sigma _{+})\tau } \\
\Sigma _{A,B,C} &=&\Sigma _{0:A,B,C}e^{(1+4\Sigma _{+})\tau } \\
\mathcal{H} &=&\mathcal{H}_0e^{2(1+\Sigma _{+})\tau } \\
\Omega  &=&\Omega _0e^{(4+8\Sigma _{+}-3\gamma )\tau }
\end{eqnarray}
This solution corresponds to a stable fixed point, but is not an exact
solution because $\mathcal{H}\equiv 0$ implies $H\equiv 0$. The fixed point
correspond to a non general-relativistic model, with the brane tension zero.
Writing out the asymptotic form of the metric in the past 
\begin{eqnarray}
ds^2=-dt^2+t^{-\frac 1{68}(\sqrt{3345}-43)}dx^2+t^{\frac 12}(dy^2+dz^2)
\end{eqnarray}
The magnetic field diverges like 
\begin{eqnarray}
h^2=\frac{3H}{1+q}t^{-1}
\end{eqnarray}
where $q$ is given as above, and $H=\frac 1{44}\sqrt{2305+44\sqrt{3345}}$.
Thus in this particular model, there are no free parameters prescribing the
metric or the magnetic field.
\begin{figure}[tbp]
\centering
\epsfig{figure=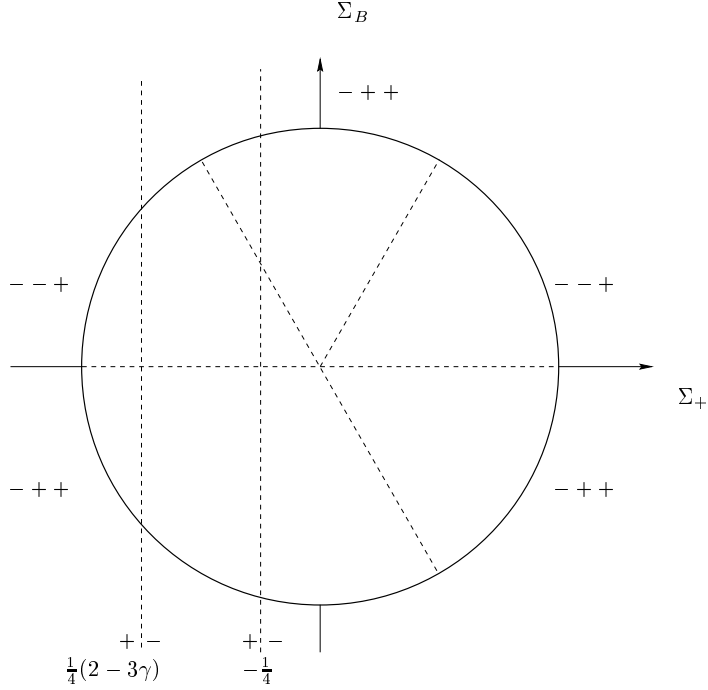, height=10cm}
\caption{The stability of the Kasner circle. The positive signs mean 
stability in the past for the different eigenvalues.}
\label{kasner}
\end{figure}

\subsection{(In)Stability of the Kasner circle}

The (in)stability of the Kasner circle proved to be an important issue in
the general relativistic case \cite{LeBlanc}. We have already noticed that
the Kasner solutions are unstable for $\Sigma _{+}>-\frac 14$. But we can
show that all points on the Kasner circle are saddle points. In our
derivations we have put $\Sigma _{A,C}=0$. Linearising the equations for $%
\Sigma _{A,C}$ around the Kasner solutions yields 
\begin{eqnarray}
\Sigma _A^{\prime } &=&(3\Sigma _{+}+\sqrt{3}\Sigma _B)\Sigma _A, \\
\Sigma _C^{\prime } &=&-2\sqrt{3}\Sigma _B\Sigma _C.
\end{eqnarray}

Thus from this alone we can say that \textit{each point on the Kasner circle
is a saddle point for past evolution}. Therefore the generic solution does
not  ``start'' at the Kasner solution. The stability analysis of the Kasner
circle is illustrated in Figure \ref{kasner}. The two horizontal lines arise
from the $\mathcal{H}^2\Theta ^2\Omega $ and the $\mathcal{H}^4\Theta ^2$
terms.

\section{Equilibrium points and stability}

Due to the complexity and the non-linearity of the equations the we will
introduce the following new variables: 
\begin{eqnarray}
H &=&\sqrt{\chi }\mathcal{H}^2\Theta  \\
\omega  &=&\sqrt{\chi }\Omega \Theta  \\
V &=&18\chi U \\
R &=&\mathcal{H}^2
\end{eqnarray}
but we will still keep the variable $\Omega $. The cost is that we get
another constraint: 
\begin{eqnarray}
H\Omega =\omega R  \label{cons2}
\end{eqnarray}
The variable $V$ can be in principle be obtained from the constraint
equation. The set of equations now becomes simpler for stability analysis,
but we see that part of the boundary of $\mathcal{M}\equiv \left\{
(H,R,\omega ,\Omega ,\Sigma _{+},\Sigma _{A,B,C})|H\Omega =\omega R\quad
and\quad H,R,\omega ,\Omega \geq 0\right\} $ has to be excluded. The
boundary is important, however, because many of the equilibrium points lie
on the boundary. We will therefore include all of the boundary in the
analysis, but bear in mind that part of the boundary is unphysical.

We must also note that the hypersurface given by (\ref{cons2}) has a conical
singularity at the origin. The tangent space at the origin therefore has a
degeneracy, (in this case is a $U(1)\times U(1)$ degeneracy). This must be
taken into account when performing a perturbation analysis around these
singular points, but it is advantageous that the singular space is an
invariant subspace of the dynamical system. Hence, through every point $p\in 
\mathcal{M}$ there is a unique maximally-extended orbit, passing through $p$.

We consider the set $\mathcal{M}$ parametrised by $(H,R,\omega ,\Omega
,\Sigma _{+},\Sigma _{A,B,C})$. The general-relativistic case is defined by
the invariant subset $H=\omega =V=0$, but if $H=\omega =0$ and $V>0$, then $%
V $ can be interpreted as a radiation fluid in the general relativity case
(note that in general $V$ can be negative). Let us examine the interesting
equilibrium points. All of those with $\omega =H=V=0$ are
general-relativistic equilibrium points and were previously found in ref.%
\cite{LeBlanc}. We check whether their stability properties change after the
inclusion of the ``non-general relativistic'' variables $\omega ,~H$ and $V$%
. We will also carry out a stability analysis for the new equilibrium
points. The results are as follows:

\begin{enumerate}
\item  $P(I)$, flat Friedmann universe \\ $R=\Sigma _{+}=\Sigma
_{A,B,C}=H=\omega =V=0$.\\ $q=\frac 12(3\gamma -2)$, $\Omega =1$, $0\leq
\gamma \leq 2$.\\ This is a future attractor in the general relativistic
models when $0<\gamma \leq \frac 43$. It is also future stable in the $H$
and $\omega $ direction for $0<\gamma <\frac 43$ and thus unstable to the
past.

\item  $\mathcal{L}R(I)$ flat Friedmann universe with radiation fluid. \\ $%
R=\Sigma _{+}=\Sigma _{A,B,C}=H=\omega =0$\\ $q=1$, $\Omega +V=1$, $\gamma
=\frac 43$, $\Omega \geq 0$.\\ This is a line bifurcation which is a future
attractor. This set of points connects the $P(I)$ and $PV(I)$ equilibrium
points.

\item  $PV(I)$, flat Friedmann universe with $V$ fluid.\\ $R=\Sigma
_{+}=\Sigma _{A,B,C}=H=\omega =\Omega =0$\\ $q=1$, $V=1$, $0\leq \gamma \leq
2$.\\ This equilibrium point is similar to that of a radiation-dominated
universe, and is stable in the future for $\gamma >\frac 43$.

\item  $PQ_1(I)$ \\ $R=\Sigma _{+}=\Sigma _{A,B,C}=H=\Omega =V=0$ \\ $%
q=3\gamma -1$, $\omega =\sqrt{6}$, $0\leq \gamma \leq 2$\\ This model is an
attractor in the past for $\gamma >\frac 43$. In the shear variables the
model is a past attractor for $\gamma >1$, but the magnetic variable $H$ is
not stable in the past for $\gamma <\frac 43$. Thus, for the generic
solutions, this is a saddle point in the past for $\gamma <\frac 43$.

\item  $\mathcal{L}Q_2(I)$\\ $R=\Sigma _{+}=\Sigma _{A,B,C}=H=\Omega =0$ \\ $%
q=1$, $\gamma =\frac 23$, $\frac{\omega ^2}6+V=1$\\ This is a one-parameter
bifurcation. Each point on it is a saddle point. It serves as a
one-parameter set of stable points to the future for $\Omega \equiv 0$.
However, this is not physically possible in the original model, since $%
\Omega \equiv 0$ implies $\omega =0$.

\item  $\mathcal{L}(I)$, flat Friedmann universe with vacuum fluid\\ $%
R=\Sigma _{+}=\Sigma _{A,B,C}=H=V=0$\\ $q=-1$, $\gamma =0$, $\frac{\omega ^2}%
6+\Omega =1$\\ This is a one-parameter bifurcation that is an attractor for
future solutions.

\item  $PM_1(I)$\\ $\Sigma _{A,B,C}=\omega =H=V=0$\\ $R=\frac 18(2-\gamma
)(3\gamma -4)$, $\Sigma _{+}=\frac 14(3\gamma -4)$,\\ $q=\frac 12(3\gamma -2)
$, $\Omega =\frac 38(4-\gamma )$, $\frac 43<\gamma <2$\\ In the general
relativistic case this is an attractor in the future for $\frac 43<\gamma
<\frac 85$, but becomes a saddle point in the brane-world model because it
is unstable in the $V$-direction for $\gamma >\frac 43$.

\item  $PM_2(I)$ \\ $\Sigma _C=\omega =H=V=0$\\ $R=\frac 14(2-\gamma
)(9\gamma -14)$, $\Sigma _{+}=\frac 14(3\gamma -4)$, $\Sigma _A=\sqrt{\frac
38(2-\gamma )(5\gamma -8)}$,\\ $\Sigma _B=-\frac 14\sqrt{3}(5\gamma -8)$, \\ 
$q=\frac 12(3\gamma -2)$, $\Omega =\frac 94(2-\gamma )$, $\frac 85<\gamma <2$
\\ In the general relativistic case this is an attractor in the future for $%
\frac 85<\gamma <\frac 53$, but becomes a saddle point in the brane-world
model because it is unstable in the $V$-direction for $\gamma >\frac 43$.

\item  $PM_3(I)$ \\ $\omega =H=\Omega =V=0$\\ $R=\frac 13$, $\Sigma
_{+}=\frac 14$, $\Sigma _A=\sqrt{\frac 7{24}}$, $\Sigma _B=-\frac 1{4\sqrt{3}%
}$, $\Sigma _C=\frac 1{2\sqrt{3}}$,\\ $q=\frac 32$, $0\leq \gamma \leq 2$\\ %
In the general relativistic case this is an attractor in the future for $%
\frac 53<\gamma <2$, but becomes a saddle point in the brane-world model
because it is unstable in the $V$-direction.

\item  $\mathcal{K}$, Kasner vacuum \\ $R=\Sigma _{A,C}=\Omega =\omega =H=V=0
$\\ $\Sigma _{+}^2+\Sigma _B^2=1$, $q=2$, $0\leq \gamma \leq 2$\\ As we have
seen, the Kasner circle is a one-parameter set of points, all are saddle
points for the generic brane-world solutions.

\item  $\mathcal{S}\mathcal{K}$, Kasner-dust half-sphere \\ $R=\Sigma
_{A,C}=\Omega =H=V=0$ \\ $\Sigma _{+}^2+\Sigma _B^2+\frac{\omega ^2}6=1$, $%
q=2$, $\gamma =1$\\ This is a two-parameter bifurcation set. The whole set
is a set of saddle points for the system of equations.

\item  $\mathcal{L}M(I)$\\ $H=\omega =V=0$\\ $R=\Sigma _A^2+\frac 1{24}$, $%
\Sigma _{+}=\frac 14$, $\Sigma _B=-\frac 1{12}\sqrt{3}$, $\Sigma _C^2=\frac
12\Sigma _A^2-\frac 1{48}$, $\frac 1{24}<\Sigma _A^2<\frac 7{24}$, $q=\frac
32$, $\Omega =\frac 78-3\Sigma _A^2$, $\gamma =\frac 53$\\ In the general
relativistic case this is a one-parameter line-bifurcation that attracts
future solutions, but in the brane-world model the points are unstable in
the $V$ direction.

\item  $PH_1(I)$ \\ $R=\omega =\Omega =\Sigma _{A,B,C}=0$ \\ $\Sigma _{+}=%
\frac{23+\sqrt{3345}}{88}$, $H^2=\frac 12\Sigma _{+}(1+4\Sigma _{+})$, $%
V=\frac 1{7744}(13725+107\sqrt{3345})$, $q=\frac 1{22}(89+\sqrt{3345})$, $%
0\leq \gamma \leq 2$\\ This is a new and unusual fixed point, and is an
attractor to the past.

\item  $PH_2(I)$ \\ $R=\omega =\Omega =\Sigma _{A,B,C}=0$ \\ $\Sigma _{+}=%
\frac{23-\sqrt{3345}}{88}$, $H^2=\frac 12\Sigma _{+}(1+4\Sigma _{+})$, $%
V=\frac 1{7744}(13725-107\sqrt{3345})$, $q=\frac 1{22}(89-\sqrt{3345})$, $%
0\leq \gamma \leq 2$\\ This is similar to the fixed point $PM_1(I)$ but it
is a saddle point.

\item  $PN_1(I)$ \\ $R=\Omega =\Sigma _{A,B,C}=0$ \\ $\Sigma _{+}=\frac
14(3\gamma -4)$, $q=3\gamma -1$\\ $H^2=\frac 1{2a}\left( -b-\sqrt{b^2-4ac}%
\right) $, $\omega =\frac{\frac 32(\gamma -1)(3\gamma -4)-4H^2}{H(3\gamma -2)%
}$ \\ $\gamma \in [\frac 1{66}(111-\sqrt{3345}),1]\cup [\frac 43,2]$\\ where 
$a=45\gamma -22$, $b=3\left( \frac{81}4\gamma ^3-114\gamma ^2+146\gamma
-56\right) $and ${c=\frac 92(\gamma -1)^2(3\gamma -4)^2}$. The local
behaviour around this fixed-point turns out to be quite complex, but a
careful analysis shows that this fixed point is saddle point for all allowed
values of $\gamma $. For $\gamma =\frac 1{66}(111-\sqrt{3345}),1$ and $4/3$, 
$H=0$ and will coincide with one of the points of $PH_2(I),\mathcal{S}%
\mathcal{K}$ and $PQ_1(I)$ respectively.

\item  $PN_2(I)$ \\ $R=\Omega =\Sigma _{A,B,C}=0$ \\ $\Sigma _{+}=\frac
14(3\gamma -4)$, $q=3\gamma -1$\\ $H^2=\frac 1{2a}\left( -b+\sqrt{b^2-4ac}%
\right) $, $\omega =\frac{\frac 32(\gamma -1)(3\gamma -4)-4H^2}{H(3\gamma -2)%
}$ \\ $\gamma \in <\frac{22}{45},\frac 23>$\\ where $a,b$ and $c$ are given
in $PN_1(I)$. This point is a saddle point for all $\gamma $.
\end{enumerate}

\begin{figure}[tbp]
\centering
\epsfig{figure=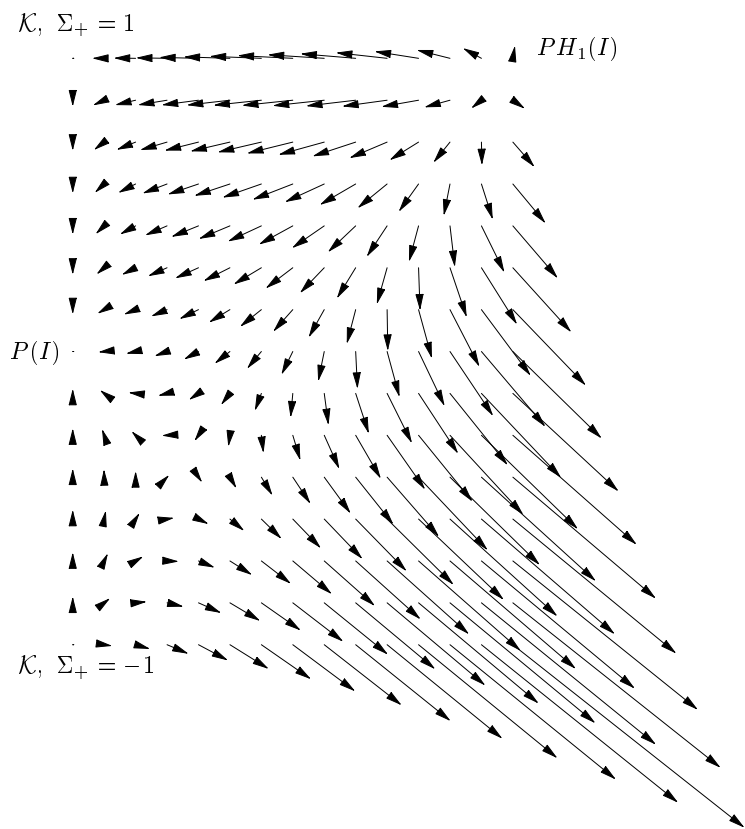, width=10cm}
\caption{A flow-diagram for the non-general-relativistic $H-\Sigma _{+}$
plane where $R=\omega =\Omega =\Sigma _{A,B,C}=0$. Five equilibrium points
can be seen; four are marked on the figure. The last one $PH_2(I)$, is a
saddle point and can be seen to the left and below the centre of the figure.}
\label{flowdiagram}
\end{figure}

\begin{figure}[tbp]
\centering
\epsfig{figure=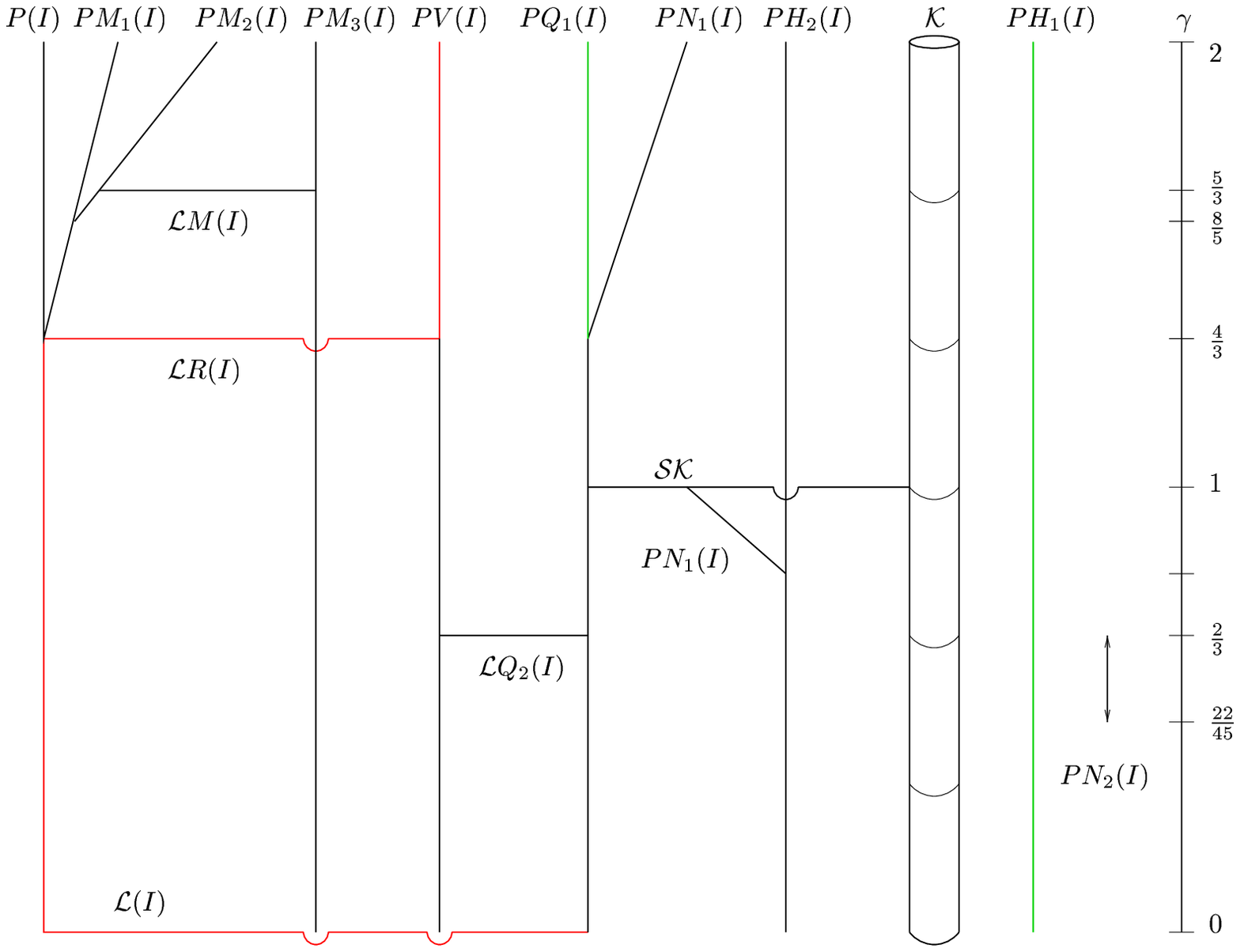, width=12cm}
\caption{A diagram of the web of equilibrium points. Horizontal lines are
bifurcation sets; all are line bifurcations except for $\mathcal{S}\mathcal{%
K,}$ which is a half-sphere. Lines in red are future attractors, while lines
in green are past attractors.}
\label{web}
\end{figure}

We can note one particular thing, the properties of the future attractor
points are altered from the general relativistic case. In the brane-world
model the dark-radiation term will make the general-relativistic future
attractors unstable for $\gamma >\frac 43$. Thus, both the local and global
properties of the phase space are altered on passing from the general
relativistic to the brane-world model. For example, we have brane-world
solutions that fly off to infinity in finite time while in the general
relativistic case the phase space is compact. Also, the brane-world
solutions behave differently in the past. The general relativistic case had
no attractors in the past for any $\gamma $. But as we can see, in the
brane-world model, there are two past attractors for $\gamma >\frac
43$ ($PQ_1(I)$ and $PH_1(I)$) and
one for $\gamma \leq \frac 43$ ($PH_1(I)$).

The importance of these fixed points can be expressed in the case $\gamma
>\frac 43$ by the following theorem:

\paragraph{Theorem}

\textsl{Let $p$ be a point in the configuration space $\mathcal{M}$
satisfying the constraint (\ref{cons2}), and let $s_p(\tau)$ be the
maximally extended solution satisfying the equations of motion such that $%
s_p(0)=p$. Then for $\gamma>\frac{4}{3}$ there exists a compact region $%
C\subset \mathcal{M}$ and a $\tau_0$ such that for every $t<\tau_0$ then $%
s_p(t)\in C$.} \\
\paragraph{Proof:} Consider the compact region $\mc{F}\subset \mc{M}$ given by
$\Sigma=\Sigma_+^2+\Sigma_A^2+\Sigma_B^2+\Sigma_C^2\leq \Sigma^*>1$,
$\Omega\leq\Omega^*>0$, $\omega\leq \omega^*>\sqrt{6}$, $R\leq R^*>0$,
$H\leq H^*>\sqrt{\frac{8}{3}}$ and assume that $\gamma>\frac 43$. For
$\Sigma> 1$ we have $\Sigma>|\Sigma_+|$ and
\beq
\Sigma'&=& 2(q-2)\Sigma+6R\Sigma_+-\left[(3\gamma-2)\omega
H+4H^2\right]\Sigma_+ \nonumber \\
&\geq & 2(\Sigma-1)\Sigma+6R\Sigma_++\Sigma\left[\frac 12 H^2-\frac 12
(3\gamma-4)\omega H +\frac 13 \omega^2\right] \nonumber \\
&\geq & 2(\Sigma-1-6R)\Sigma
\eeq
because the polynomial in $\omega$ and $H$ inside the square brackets
is positive definite for $\gamma\in[\frac 43,2]$. 
Similarly, for $H\geq H^*$ we have
\beq
H'\geq 12\left(H^*-\sqrt{\frac{8}{3}}\right)
\eeq
and for $\omega\geq \omega^*$ we have
\beq
\omega'\geq (3\gamma-2)\left(\frac{(\omega^*)^2}{6}-1\right)\omega^*
\eeq
These equations ensures that the variables $H$ and $\omega$
are finite in the past. We also have for $H\neq 0$
\beq
\left(\frac{R}{H}\right)'=(1+q)\left(\frac{R}{H}\right)\geq 2
\left(\frac{R}{H}\right)
\eeq
Integrating yields the bound $\frac{R(\tau)}{H(\tau)}\leq
\frac{R(0)}{H(0)}e^{2\tau}$ for $\tau<0$. Hence, in the past 
$R$ can be driven arbitrarily close to zero as $\tau\lra
-\infty$ since $H$ is bounded. Thus for $\Sigma>\Sigma^*$, $\Sigma'>0$
for some time in the past. Hence, $\Sigma$ is bounded in the past too. 
Similarly, for $\omega\neq 0$, $\Omega$ can also be driven
arbitrarily close to zero. Thus in this case, the region $\mc{F}$ is
sufficient and we can set $C=\mc{F}$. 

If $H=0$ (which is an invariant solution subspace) the
equation for $V$ is
\beq
V'=(2q-2)V
\eeq
which after a rearranging yields
\beq
(\ln |V|)'=2q-2\geq 0
\eeq
Thus $V$ is bounded in the past, and according to the constraint
equation all of the other variables have to be as well. 

If $\omega=0$ but $H>0$, then $\Omega'\geq
(3\gamma-4)(\Omega^*-1)\Omega^*>0$ for $\Omega>\Omega^*>1$ which makes
$\Omega$ bounded in the past. 

The theorem
now follows from this analysis. $\blacksquare$ \\

This theorem shows that the solution
cannot fly off to infinity as $\tau \longrightarrow -\infty $. With some
amendments, we believe that the behaviour at the initial singularity is
``well-behaved'' for any $\gamma $, even though we have not been able to be
prove a similar theorem in that case.

In Figure \ref{flowdiagram} we have drawn a flow-diagram of the dynamical
system. The section shown is a part of the non-physical boundary, and the
vector field is projected onto this plane. In Figure \ref{web} the web of
equilibrium points are illustrated. We see how the points are connected
through bifurcation sets, and it is clear how some branch off from other
equilibrium points.

From this analysis we believe that it is very likely that the pure magnetic
Bianchi type I brane-worlds do not exhibit chaotic behaviour in the past.

\section{Discussion}

In this paper we have investigated pure magnetic Bianchi type I
brane-worlds. We found asymptotic forms for the solutions near the initial
singularity for different configurations of matter, magnetic field and shear
anisotropy.. We also checked whether there are any equilibrium points that
could serve as a past attractor for the generic solutions. For $\gamma
>\frac 43$ we found two such attractors, while in the case $\gamma \leq
\frac 43$ there was one. This may indicate that there can be no chaotic
behaviour in the generic solutions for a magnetic Bianchi type I
brane-world. To the past, the solution can approach the general relativistic
case but will only oscillate a finite number of times. The possible initial
states of these two equilibrium points, correspond to two different non
general-relativistic models, one of them appears to be previously unknown.

When we investigated the various equilibrium points, we noted that all of
the future stable general-relativistic points were unstable in the
brane-world scenario for $\gamma >\frac 43$. Another major difference
between the two cases is that
the configuration space is non-compact for the brane-world solutions. We
found solutions that fled to infinity in a finite coordinate time, and found
that the non-local energy $\mathcal{U}$ can become arbitrary large (and even
negative) and so presents a major obstacle to this sort of analysis.

The models examined in this paper are based upon the simplest anisotropic
universes. It remains to be seen what new features might emerge from a study
of the general case. However, we know that in the magnetic case (in contrast
to the non-magnetic situation) many of the features of the general
anisotropic universe are present in the simple type I model.
\paragraph{Note:} After this paper was completed and submitted, a
similar and independent work by Alan Coley appeared \cite{Coley1,Coley2}
addressing some of the same issues as we have done in this paper.

\section*{Acknowledgments}

SH was funded by The Research Council of Norway.

\end{document}